\documentstyle[11pt,nyas,psfig]{article}

\begin{document}

\newcommand\th{\thinspace}
\newcommand{\beq}{\begin{equation}}
\newcommand{\greq}{\begin{equation}\left\{ \begin{array}{l}}
\newcommand{\egreq}{\end{array}\right. \end{equation}}
\newcommand{\egreqn}[1]{\end{array}\right. \label{#1}\end{equation}}
\newcommand{\eeq}{\end{equation}} 
\newcommand{\eeqn}[1]{\label{#1}\end{equation}} 
\newcommand{\noi}{ \noindent }
\newcommand{\lp}{ \left(}
\newcommand{\rp}{ \right)}
\newcommand{\lc}{ \left[}
\newcommand{\rc}{ \right]}
\newcommand{\disp}[1]{\displaystyle #1}
\newcommand{\dnz}[1]{\frac{d  #1}{dz}}
\newcommand{\dz}[1]{\frac{\partial  #1}{\partial z}}

\title{Plumes in stellar convection zones}

\author{J.-P. Zahn}
\affil{D\'epartement d'Astrophysique Stellaire et Galactique, Observatoire de Paris, 
Section Meudon, 92195 Meudon, France}

\begin{abstract}
All numerical simulations of compressible convection 
reveal the presence of strong downwards directed flows. Thanks to
helioseismology, such plumes have now been detected also at the top of the solar
convection zone, on supergranular scales. Their properties may be crudely described
by adopting Taylor's turbulent entrainment hypothesis, whose validity is well
established under various conditions. Using this model, one finds that the strong
density stratification does not prevent the plumes from traversing the whole
convection zone, and that they carry upwards a net energy flux. They penetrate to
some extent in the adjacent stable region, where they establish an almost adiabatic
stratification when there is little radiative diffusion. These plumes have a strong
impact on the dynamics of stellar convection zones, and they play probably a key role
in the dynamo mechanism.    

\bigskip

{\small \bf \noi Proceedings of the Fourteenth International Annual 
Florida Workshop in Nonlinear Astronomy and Physics,
``Astrophysical Turbulence and Convection''\th\th, 
University of Florida, Feb. 1999

\noi (to appear in the Annals of the New York Academy of Sciences)}

\end{abstract}


\keywords
{Convection, hydrodynamics, turbulence; solar and stellar interiors.}

\vspace{-0.5cm}

\section{Introduction}
One of the main weaknesses of stellar physics remains our poor description
of thermal convection. It is true that the 
widely used mixing-length treatment permits to
construct models which represent fairly well the gross properties of stars, 
but it fails when 
one attempts to apply it to more subtle processes, such as convective
penetration, the differential rotation of the 
Sun, or its magnetic activity. 

The situation is rapidly changing however. Over the two past decades significant
progress has been achieved through numerical simulations of
increasingly `turbulent' convection in a stratified medium. 
These have shown that fully compressible
convection is highly intermittent, displaying strong, long-lived, 
downwards directed flows, which contrast with the slower, random upward motions
(Hurlburt et al. 1986; Cattaneo et al. 1991;
Nordlund et al. 1992; Muthsam et al. 1995; Singh et al. 1995; Brummel et al. 1996;
Stein \& Nordlund 1998). These coherent structures are called {\it plumes}
(`panaches' in french), by analogy with those observed in the Earth atmosphere. They
originate in the upper boundary layer, where they are initiated by the strong
temperature and density fluctuations, which arise there in the steep superadiabatic
gradient. 

True, these numerical results are still
difficult to apply as such to the
convection zone of a star, because of
the huge gap in the relevant control parameters,
the Reynolds and the Prandtl numbers. 
But they suggest that
coherent structures may play an important role in the dynamics of a
turbulent convective layer, and thus open the possibility of a different
description of stellar convection. 

\begin{figure}
\centerline{\psfig{figure=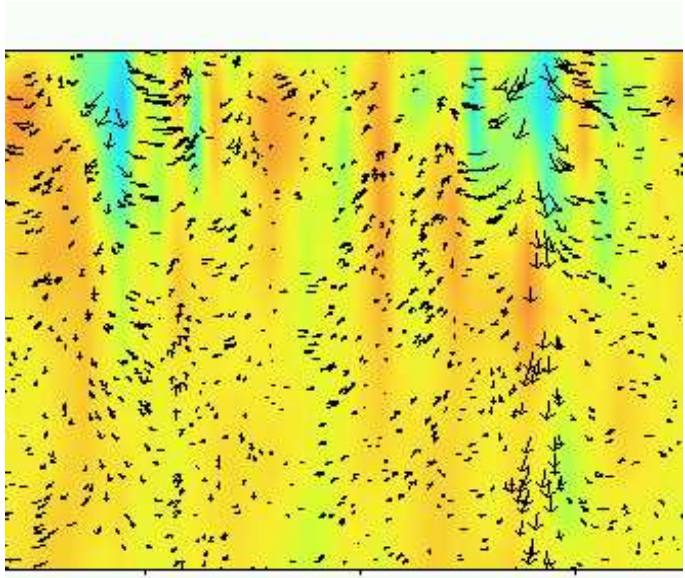,width=9cm}}
\caption{\small Time-distance tomography with the MDI instrument on board of~SOHO; 
the field shown measures 8 Mm in depth and 150 Mm in width.
(http://sohowww.nascom.nasa.gov/gallery/MDI/mdi009.gif)}
\label{tomofig}
\end{figure}

And indeed these plumes have now been detected in the upper layers of the solar
convection zone, thanks to time-distance tomography. This powerful helioseismic
technique reveals the presence of downdrafts apparently associated with the
supergranular network, in which the temperature is strongly correlated with the
vertical velocity (D'Silva et al. 1996; Duvall et al. 1997). Fig. \ref{tomofig}
shows an example of such tomographies; it can been downloaded from the SOHO gallery
on the Web.


Turbulent plumes are commonly
observed in the Earth atmosphere, where they rise above concentrated heat sources 
(chimneys, nuclear plants, 
volcanos, etc.). Their theoretical interpretation is still based on the entrainment  
hypothesis,
which was first proposed by G.I. Taylor (cf. Morton et al. 1956); we recall it
below.  In astrophysics, plumes have been invoked by Schmitt et
al. (1984) as responsible for the penetration into the stable domain
beneath the solar convection zone. Somewhat later an estimate of that
penetration has been given by Zahn (1991), which is also
based on a crude plume model.

More recently, Rieutord and Zahn (1995) moved a step forwards, by letting the plumes
develop through the whole convection zone. They adapted to a medium which
is highly stratified, in density and temperature, the
treatment which is used in geophysical fluid dynamics. Furthermore, 
they took into account the backflow which is induced by the confinement
of the plumes in a limited volume, and which determines the depth the plumes may
reach.  This review recalls the main properties of that model; it will also address
the role the plumes may play in convective penetration and in the generation of gravity waves.

\section{One single plume in an isentropic stratification}
We shall first examine the basic properties of a turbulent plume
which resides in an isentropic, plane-parallel atmosphere, assuming
that the plume is fed steadily by a cold layer of fluid lying on the top 
of the atmosphere.
A schematic view
of the flow in such a plume is given in fig.~\ref{plume}.


\begin{figure}
\centerline{\psfig{figure=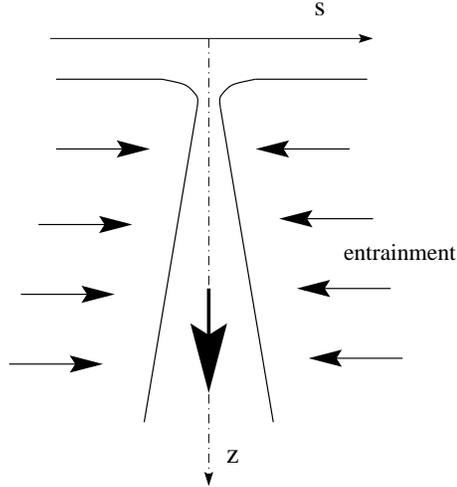,width=6cm}}
\caption{\small The plume flow in an adiabatic atmosphere 
(after Rieutord \& Zahn 1995; courtesy A\&A).}
\label{plumfig}
\end{figure}

\subsection{Governing equations}
The equations governing the structure of turbulent plumes have been
established in the original work of Morton et al. (1956).
One assumes that the flow is stationary, that the plumes are axisymmetric about 
the vertical, and that all horizontal variations, 
namely those of the vertical velocity $v_z$, 
of the density excess $\delta\rho$ and of the excess of specific enthalpy 
$\delta h$, have the same gaussian profile:
\greq
\disp{v_z(r,z) = V(z)\exp(-s^2/b^2)}\\
\disp{\delta\rho(r,z) = \Delta\rho(z)\exp(-s^2/b^2)}\\
\disp{\delta h(r,z) = \Delta h(z)\exp(-s^2/b^2)}
\egreq
where $b\equiv b(z)$ is the effective radius of the plume. For
convenience, we take the vertical coordinate $z$ pointing downward;
the density
and the specific enthalpy of the isentropic atmosphere will be designated by
$\rho_0 (z)$ and $h_0(z)$.

The equations describing the vertical profile of the plume are derived
from the three basic equations of fluid mechanics expressing 
the conservation of mass, momentum and
energy.

\subsubsection{Conservation of mass.}
In the steady flow of the plume, the mass conservation implies
\[ \hbox{div}(\rho\vec{v}) = \frac{1}{s}{\partial \over \partial s} \lp{s\rho
v_s}\rp +  {\partial \over \partial z} \lp{\rho v_z}\rp =0 \ ; \] 
when integrated over $s$, this equation becomes
\beq \dnz{}\int_0^{+\infty}\!\rho v_z sds\; +\; [s\rho v_s]_0^{+\infty} = 0 .
\label{mass} \eeq

The entrainment hypothesis made by G.I. Taylor postulates that the radial
inflow of matter into the plume is proportional to the central vertical velocity:
\beq \lim_{s \rightarrow +\infty} sv_s = -\alpha b(z)|v_z(0,z)| 
\label{tayl}\eeq
where $\alpha$ is the entrainment constant. The absolute value
guarantees that the plume is always accreting matter, whether it is directed
upwards or downwards.
Using the gaussian profile of  $v_z$ and
$\delta\rho$, one casts the mass equation in its final form
\beq
\dnz {}\left[ \rho_0b^2 V \right] = 
2\alpha\rho_0 bV \, ,
\label{masse1}
\eeq
in the limit of vanishing density contrast $\Delta\rho/\rho_0 \ll 1$.

\subsubsection{Conservation of momentum.}
Ignoring the viscous stresses, the flow obeys 
the steady Euler equation:
\[ \partial_j(\rho v_jv_i + \delta p\, \delta_{ij}) = 
\delta\!\rho\, g_i \, , \]
\noi where the static equilibrium has been subtracted. We assume that
the plume is in pressure
balance with the surrounding medium; it leads us to
\[ \frac{1}{s}{\partial \over \partial s} \lp{s\rho v_s v_z}\rp + 
{\partial \over \partial z} \lp{\rho v_z^2}\rp = \delta\!\rho\, g \] 
($g$ is the gravity, which we assume constant). After integration as before, we get
\beq
\dnz{}\left[ \rho_0b^2 V^2\right] = 2gb^2\Delta\rho \, .
\label{moment1}
\eeq

\subsubsection{Conservation of energy.}
We start from the steady energy equation
\[ \hbox{div}(\rho e\vec{v})=\hbox{div}(\chi\vec{\nabla} T) - p\hbox{div}\vec{v} \ ,\]
\noi where $e$ is the specific internal energy and $\chi$ the radiative
conductivity, and rewrite it by
using the momentum equation and the definition of the specific enthalpy 
$dh \equiv de + d(p/\rho)$:
\[ \hbox{div} \left[ (h+{\frac{1}{2}} v^2)\rho\vec{v} - \chi\vec{\nabla} T \right] = 
\rho\vec{v}\cdot{\vec{g}} . \]
This equation may be further simplified by remembering that
\[ {\vec{g}} = \vec{\nabla} h_0 \]
\noi in an isentropic atmosphere (see below).
Finally, the conservation of the energy flux is expressed in conservative form:
\beq \hbox{div} \lc \lp \delta h+{\frac{1}{2}} v^2 \rp \rho\vec{v} - 
\chi\vec{\nabla} T \rc = 0  ,\eeq
where we recognize respectively the enthalpy, kinetic and radiative
fluxes.

To make contact with previous work, we first neglect the radiative flux.
Proceeding as before, we 
integrate in the radial direction, and obtain
\beq {\frac{1}{2}}  \rho_0b^2 V\Delta h +
\frac{1}{6} \rho_0b^2 V^3 =  -F/\pi \ ,
\eeqn{ener}
where $F$ designates the total flux (enthalpy plus kinetic) carried
by one plume.

From now on, we shall drop the subscript $_0$, since there will be no ambiguity
anymore between the local values of density and temperature, and their
counterparts in the isentropic atmosphere.

\subsubsection{Thermodynamics.}
For simplicity, we shall assume that the fluid is 
a perfect
monatomic gas, with $\gamma = 5/3$ being the adiabatic exponent. We thus 
complete the system of governing equations with the two thermodynamic
relations:
\[ \Delta h = c_p  \Delta T \]
\[ {\Delta \rho \over \rho} +  {\Delta T \over T} = 0  \]
\noi where the latter expresses the pressure balance between the
plume and the surrounding medium ($c_p$ is the heat capacity at constant pressure).

\subsubsection{The isentropic atmosphere.}
As is well known, the isentropic atmosphere is a polytrope of 
index $q=1/(\gamma - 1)$. In plane-parallel geometry, 
the density and the temperature vary with depth $z$ as:
\greq
\disp{\rho (z) = \rho_{i}(z/z_i)^q} \\
\\
\disp{T(z) = T_{i}(z/z_i) }.
\egreqn{poly}
The origin of the vertical coordinate $z$ is taken at the `surface', where
pressure, density and temperature all vanish, and the
subscript $_{i}$ designates the reference level for which we choose here the base
of the atmosphere. The reference temperature $T_{i}$ is related to the 
reference  depth by $z_i  = c_p T_{i}/ g$.

\subsection{Asymptotic behavior}
The equations governing the plume flow are nonlinear and in most cases
it is necessary to
integrate them numerically, with boundary conditions imposed at the start of
the plume, for instance on $\Delta \rho$, $b$ and $V$. However, the system
has asymptotic solutions which can easily be obtained in analytic form.
The one which presents most interest corresponds to
the developed phase of the downwards directed plume, when it is fully controlled
by entrainment. 
The solution then obeys the power laws
\beq b = b_i \lp {z / z_i} \rp^p, \quad V = V_i \lp {z / z_i} \rp^r,
\quad \Delta \rho = \Delta \rho_i \lp {z / z_i} \rp ^m \label{asreg}\eeq
where the exponents take the values:
\greq 
p = 1 \\
\\
\disp{r = -\frac{q+2}{3} = - \frac {2 \gamma -1} {3 (\gamma - 1)} }\\
\\
\disp{m = -\frac{2q+7}{3} = - \frac {7 \gamma - 5} {3 (\gamma - 1)} \, .}
\egreq
A direct consequence of Taylor's
entrainment hypothesis (\ref{tayl}) is  that the plume
radius increases linearly with depth. The half-angle $\beta=$atan$(b_i/z_i)$ of
the cone is related to the adiabatic exponent:
\beq 
\tan \beta = \frac{3\alpha}{q+2} = \frac{3\alpha (\gamma -1)}{2\gamma -1}
\, . \label{beta}
\eeq
With $\gamma = 5/3$ (perfect monatomic gas), we have $\tan \beta = 6\alpha/7$,
whereas $\tan \beta = 6\alpha/5$ in the absence of stratification. The
entrainment coefficient $\alpha$ has  been determined experimentally in various
situations, and the value 0.083 is widely adopted (Turner 1986); we shall assume
that this value also applies to our fully ionized gas.

For given initial conditions imposed at the top of the plume, the solutions
rapidly reach their asymptotic regime: the plumes start contracting while they
accelerate, and then they take their characteristic conical shape (cf. Rieutord
\& Zahn 1995).  

Note that one does not retrieve the
incompressible case by letting the polytropic index tend to zero, 
$q \rightarrow 0$.  The
reason for this singular limit is the assumption made in the classical
treatment (Morton et al. 1956) that the surrounding medium is isothermal,
while in our stratified atmosphere the temperature grows linearly with depth. 
Let us emphasize however that
our asymptotic solution (\ref{asreg}) is rather special, since 
its focal point $z = 0$ coincides with the `surface' of the
atmosphere.

\subsection{Turbulent plumes carry upwards a net energy flux}
An interesting property of this asymptotic regime is the strict
proportionality between the flux of kinetic energy and that of enthalpy:
\beq R = \frac{\disp{\int {\frac{1}{2}}\rho V^3 sds}}
{\disp{\int\rho c_p \, \delta T \, V sds}} 
=  cst . \eeq
With the gaussian profile we have adopted for the plume, this ratio
is given by
\beq R = - \frac{2}{q+2} = - \frac{2\gamma-2}{2\gamma -1} \; . \eeq
Its value is 1 for the Boussinesq fluid (no density stratification)
but it is 4/7 in the isentropic atmosphere (for the perfect monatomic gas), meaning
that the net energy flux is directed upwards. 

In their numerical experiment Cattaneo et al. (1991) observed to
their surprise that the strong downdrafts transported a flux of kinetic energy
which very nearly canceled the flux of enthalpy. The reason is that their
plumes were not turbulent, but laminar flows which satisfied approximately
Bernoulli's theorem. In recent simulations, which have been achieved with better
spatial resolution, there is more entrainment, and the plumes do carry a net
energy flux upwards.

\section{Towards a more realistic model; application to the solar convection zone}
Up to now, we have considered the case of a single plume
in an unlimited isentropic plane layer. To progress towards a more realistic
situation, we must take into
account the spherical geometry of the fluid layer and include the variations of
gravity with depth.
Here we shall apply the model to the solar convection zone (SCZ); the convective 
core with its rising plumes has been treated by Lo and Schatzman (1997).

We neglect the mass of the envelope, so that the
density and temperature field of the background are given by 
\greq
\disp{T(z) = T_i \; \frac{\eta}{1-\eta}\lp \frac{1}{x}-1\rp}\\
\\
\disp{\rho(z) = \rho_i \lp{T(z)\over T_i}\rp^q} \\
\\
\disp{x = 1 - z} 
\egreqn{atmos2}
where $x=r/R_\odot$ is the radial coordinate scaled by the
radius, and $\eta=r_{\rm cz}/R_\odot$ its value at the base of the SCZ (our reference
level). 
 
We shall also take into account the
radiative flux in the energy equation, thus (\ref{ener})
now reads
\beq {\frac{1}{2}}  \rho b^2 V\Delta h + \frac{1}{6}\rho b^2 V^3
  =  -(F_{\rm tot}- F_{\rm rad})/\pi \ . \label{ener2}
\eeq
With our assumption that all the
convective energy is transported by the plumes, whose number is $N$, the total
flux amounts to the luminosity of the star: 
\beq N \, F_{\rm tot} = L_\odot . \eeq 
The radiative contribution to the flux carried by one plume is $F_{\rm rad}$, where
\[ N \, F_{\rm rad} = L_{\rm  rad} =  
- 4 \pi r^2 \, \chi \lp \dz{T}\rp_{\rm ad}, \quad \chi= \frac{16\sigma
T^3}{3 \kappa \rho}\, , \] 
with  $\chi$ being the radiative
conductivity, $\kappa$ the opacity, and $\sigma$ the Stefan constant.  

\subsection{Plumes diving in a rising counterflow}
We shall now take into account that there
are numerous plumes and that they share a finite volume.  This
has two consequences: the plumes will interact and possibly merge, and they
will move in a rising counterflow. 
Let us first estimate the
maximum number of plumes which may coexist in the SCZ, assuming that they
all originate at the top and that all reach its bottom without being hindered by the
counterflow. This number is given by the ratio of the area of the base of the SCZ to
the section of a single plume  at this depth:
\[ N = 4 \left[{\eta \over (1-\eta) \tan \beta }\right]^2 \]
where $\beta$ is given by (\ref{beta}), and $\eta = r_{\rm cz}/R_\odot$ is the
scaled radius of the base of the SCZ. Taking $\eta = 0.7$, the result is
$ N_{\rm max} \sim 4300$.
We see that even if the plumes were closely packed, they would not be very 
numerous. But they will be actually less, because as soon as their occupy more
than half of the total area, at depth $z_i/ \sqrt{2}$, the velocity of 
the backflow exceeds that of the plumes, and this will affect their dynamics.

To model the effect of this counterflow,
we shall assume that the entrainment of mass into the
plume is proportional to the relative velocity of the
plume with respect to the surrounding fluid. This yields the
mass conservation equation
\beq
\dnz{}\left[ \rho b^2 V\right] = 2\alpha \,\rho b(V-V_u)
\label{masse2}\eeq
where $V_u$ is the upward velocity of the surrounding fluid; since 
$V_u<0$, entrainment is enhanced. The momentum equation needs
also be completed by a term taking into account the entrainment of
momentum into the plume. We thus transform (\ref{moment1}) into
\beq
\dnz{}\left[  \rho b^2 V^2\right] = 2gb^2\Delta\rho
+ 4 \alpha \, \rho b V_u(V-V_u) \, .
\label{moment2}
\eeq
Equations (\ref{masse2}) and (\ref{moment2}) are completed by 
\beq (4\pi r^2 -N\pi b^2)\rho V_u + N\pi b^2\rho V = 0 \label{flux0}\eeq
which expresses the global conservation of mass.
The consequence of these additional effects is that plumes are able to reach
the bottom of the convection zone only if they are strong enough, and not
too numerous. From the numerical integration of these equations
 it turns out that only about $1000$
plumes are able to reach the bottom of the SCZ (see Rieutord \& Zahn 1995).

\subsection{Plume coalescence}
At the top of the SCZ, the number of
plumes is probably of the same order as that of the granules, which is
considerably larger than 1000. Hence a drastic
reduction of the plumes number must occur as depth increases. 
The plumes are stopped at a level
which depends on their strength: only the 1000 strongest are able to reach the
bottom. The others will either dissolve in the
(turbulent) backflow and disappear as such, or they will merge with a stronger
companion.  

Such merging is observed in the  numerical simulations 
(Stein \& Nordlund 1989; Spruit et al. 1990): 
two neighboring plumes are advected by each other, since they both 
accrete surrounding
fluid, and they are pulled to each other until they finally coalesce. 
The merging depth
can be estimated by assuming the
flow be in the asymptotic regime; the result is
\beq z_m = \frac{d}{\sqrt{2\alpha\beta}} \eeqn{mdepth}
where $d$ is the initial separation of the two plumes, which we
assumed here to be identical.
Using the value of $\beta$ for the monatomic ideal gas and
$\alpha=0.083$, we get
$ z_m \sim 9.2 \, d \ .$

Similar coalescences will then repeat at increasing depths, until the survivors
reach the base of the convection zone, in a sort of inverse cascade
with a large scale flow building up from smaller scales. 

\section{Penetration at the base of a convection zone}
When the plumes reach the bottom of the unstable domain, they still possess
a finite velocity which enables them to penetrate some distance
into the stable, subadiabatic region, where they establish a nearly 
adiabatic stratification by releasing their entropy when they come to rest.
A first attempt to estimate the extent of penetration of such plumes
was made by Schmitt et al. (1984). 
However, they did not include the return flow,
and they had to impose both the velocity $V$ and the filling factor $f$ at 
the base of the unstable domain because they did not solve the plume equations
above that level. They found empirically that the penetration depth varies as
$f^{1/2} V^{3/2}$, and this scaling can be easily explained.


\begin{figure}
\centerline{\psfig{figure=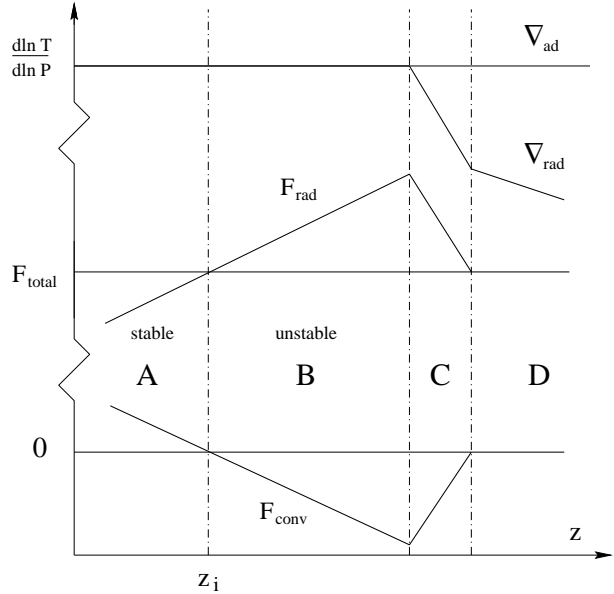,width=8cm}}
\caption{\small Schematic structure of the base of a convective envelope 
(after Zahn 1991; courtesy A\&A).}
\label{penfig}
\end{figure}

\subsection{An estimate of the penetration penetration depth}
The stratification at the base of the a convective envelope is sketched in
fig. (\ref{penfig}). (A) designates the unstable region, where the
temperature gradient is maintained close to adiabatic by the convective
motions. One thus assumes that the radiative leaks are negligible compared to the
advective transport of heat, which is the case when the P\'eclet number is much larger
than unity:
\beq
Pec = {v \ell \over K} \gg 1,  \quad K = {\chi \over \rho c_p},
\eeq
where $K$ is the thermal diffusivity and $\chi$ the radiative conductivity. 
$v$ and $\ell$ are the velocity and size characterizing the convective motions,
i.e. here the central velocity $V$ and the width $b$ of our plumes. At the base
of the SCZ, $Pec \approx 10^6$.
 
Due to the steady increase
with depth of the radiative conductivity $\chi$, the radiative flux 
$F_{\rm rad}=\chi (dT/dz)_{\rm ad}$ rises until it equals the total flux 
$F_{\rm total}$ at the level $z=z_i$, where also the radiative gradient equals the
adiabatic gradient:  $(dT/dz)_{\rm rad} = (dT/dz)_{\rm ad}$. If there were
no convective penetration, this would be the edge of the convection zone, as predicted
by the Schwarzschild criterion, and the temperature gradient would thereafter decrease
as $dT/dz = F_{\rm total}/\chi$. 
But the motions penetrate into the stable region (B) and they
render it nearly adiabatic over some distance $d_{\rm pen}$, while being decelerated by
the buoyancy force.  When the P\'eclet number has dropped below unity, the temperature
gradient settles from adiabatic to radiative in a thermal boundary layer (C).

Since the penetration depth is rather small, one may
simplify the problem by ignoring the variation with depth of most quantities
(density, width of the plume, etc.), and by neglecting the kinetic energy flux
and the turbulent entrainment (or detrainment). One keeps of course the
variation with $z$ of the conductivity; in the vicinity of $z_i$,
the radiative flux is approximated by
\beq
F_{\rm rad} = \chi \lp{dT \over dz}\rp_{\rm ad} = F_{\rm total} \left[ 1 + \lp{d \chi
\over dz}\rp_i  \lp z-z_i\rp \right] \, .
\label{fluxrad}
\eeq
Therefore the convective flux varies as
\beq
F_{\rm conv} = - F_{\rm total}  \lp{d \chi \over dz}\rp_i 
\lp z-z_i\rp  \, .
\label{fluxconv}
\eeq
This enthalpy flux may be expressed in terms of the vertical velocity and of
the temperature contrast in the plumes, as in (\ref{ener}):
\beq
F_{\rm conv} = - f \rho c_p V \Delta T \, ,
\label{fluxconv1}
\eeq 
where we have introduced the filling factor $f$ of the plumes, defined as the
fractional area covered by them: $f = N b^2 /R^2_{\rm cz}$.

To estimate the penetration depth, we follow the plumes from $z=z_i$, where
their velocity is $V_i$, until they stop at 
$z = z_i + d_{\rm pen}$. Their deceleration is described to first order by
\beq
{1 \over 2} {d V^2 \over dz} = g {\Delta \rho \over \rho} = 
- g {\Delta T \over T} .
\label{decel}
\eeq
After elimination of $\Delta T$ with (\ref{fluxconv1}), the integration of
(\ref{decel}) yields the following expression for the penetration depth
$d_{\rm pen}$:
\beq
d_{\rm pen}^2 = {3 \over 5} H_p H_\chi \left\{f {\rho V_i^3 \over F_{\rm total}}
\right\} = {3 \over 5} H_p H_\chi Q \,
, \label{pendepth}
\eeq
where $H_p$ is the scale-height of the pressure and $H_\chi$ that of the
radiative conductivity. This is the scaling obtained empirically by Schmitt et
al. (1984).

At that point one may argue that the term $Q$ in curly brackets is determined by
the dynamics in the convection zone, and that it will not depend much on its
depth, provided this depth is large enough and that convection is sufficiently
adiabatic. $Q$ is probably smaller than unity: in the mixing length treatment,
$Q = (1/10) \, \Lambda / H_p$ in the bulk of the convection zone. This
figure, with $\Lambda / H_p \approx 1.5$ and $H_\chi \approx H_p/2$, would yield a
penetration depth of the order of $1/5$ of the pressure scale-height at the base of
the solar convection zone.

\subsection{Penetration of plumes originating at the top of the convection zone}
Rieutord and Zahn (1995) computed the penetration in a more consistent way, 
by integrating the governing equations from the top of
the convection zone until the plumes come 
to rest. When neglecting the backflow, the penetration extends to about one half
of the pressure scale-height, independent on the number of plumes. But the picture
changes drastically when one includes the interaction of the plumes with the
backflow. Then the penetration depth depends on the number of plumes: it decreases
almost linearly with $N$, from one pressure scale-height for $N \rightarrow 0$ to $0$
for $N \approx 1200$. The reason for this trend is that the backflow increases with
the number of plumes, and that it slows down their motion by loading them with
upward momentum. 

Unfortunately, this model is unable to predict the number of plumes. Moreover, a much
more sophisticated model is needed to describe the termination of the plume flow,
which would include  the loss of mass by the plume in this region, or
`detrainment', and its transfer to the upward motion. 

\subsection{Penetration vs. overshooting}
What we have called convective {\it penetration} is that part of the excursion of
convective motions into the stable region below where they enforce an almost
adiabatic stratification ($\nabla_{\rm ad} - \nabla \ll 1$). It corresponds to
region (B) in fig. \ref{penfig}, in which the P\'eclet number is substantially
larger than unity (we recall that at the base of the SCZ, $Pec \approx 10^6$). 
In comparison, the thermal adjustment layer (C) is very thin: about 1 km in the
Sun, assuming that all plumes have the same strength (Zahn 1991).

The situation is very different in stars which possess a shallow convection zone.
For instance, at the base of the convection zone of an A-type star, $Pec \approx
1/100$, which means that the convective motions which enter the stable layer below
rapidly settle in thermal equilibrium with their surroundings, and that
they hardly feel the buoyancy force (Toomre et al. 1976; Freytag et al. 1996).
The region (B) has disappeared -- all what remains is region (C). To insist on the
contrast with the almost adiabatic penetration at high P\'eclet number, we prefer to
call this convective {\it overshoot} (or undershoot).

This is not only semantic matter. To take one example, numerical simulations which
are performed to represent A-type stars cannot be used to predict the amount of 
penetration below the solar convection zone, as was attempted by Bl\"ocker et al.
(1998). In fact, for lack of spatial resolution, present day simulations are unable
to achieve a P\'eclet number which would realistically describe convective
penetration; for instance, in the calculations made by Hurlburt et al. (1994),
regions (B) and (C) are of comparable size, and a substantial fraction of the
extension into the stable region is due to overshoot.

\subsection{Generation of internal gravity waves}
In the laboratory, convective penetration is observed to generate internal
waves, which propagate in the stable adjacent layer (Townsend 1966; Adrian 1975). the
same is also expected at the boundary of stellar convection zones. Such waves have
indeed been observed in numerical simulations of penetrative convection, 
where they seem to be produced by the pronounced downdrafts (Hurlburt et al.
1986; Hurlburt et al. 1994), which are the 2-dimensional analogues of
3-dimensional plumes. In these calculations one observes a strong feedback of
the waves on the downdrafts, because the waves are reflected back by the lower
boundary of the computational domain, a property which is not expected from the
solar gravity waves produced by the plumes.

The generation of gravity waves by turbulent plumes is presently being
investigated by M. Kiraga, in 2- and 3-dimensional simulations. He takes great care
of avoiding the reflexion of the waves by the lower boundary of his computational
domain, by implementing in its vicinity a strong viscous damping. The first results
have been presented at the Granada workshop on convection (Kiraga et al. 1999); they
look promising, and will be used to test the prescriptions which have been proposed
for estimating the flux of gravity waves.

Our interest in these gravity waves is motivated by the fact that they may extract
angular momentum from the radiative interior of solar-type stars (see Kumar \&
Quataert 1997; Zahn et al. 1997; Kumar et al. 1999). 

\section{Summarizing the properties of turbulent plumes}
There is little doubt that the convection zone of solar-like stars exhibits 
strong spatial intermittency, with downwards directed plumes carrying a
substantial fraction of the energy flux. Such plumes are seen in all numerical
simulations, and they have now been observed in the uppermost part of the solar
convection zone. 

If Taylor's parametrization of the entrainment of surrounding fluid into
turbulent plumes holds in the hot plasma of stellar interiors, and we see no
reason why it should not, then we expect such plumes to
traverse the whole convection zone. 
According to the crude model presented above in \S2, these plumes behave much as if
they were  traversing an isentropic atmosphere, with no feed-back at all, and
ignoring the radiation flux. 
Shortly after their start,
they reach an asymptotic regime with their size increasing linearly with
depth. Due to the density stratification, the cone angle $\beta$ is somewhat smaller
than in the Boussinesq case treated so far: $\tan \beta =6\alpha /7$ instead
of $6\alpha /5$, with $\alpha$ being the entrainment constant. 

These plumes carry kinetic energy downwards, i.e. in the `wrong'
direction. But their enthalpy flux always exceeds
the kinetic flux: the ratio between the two is constant in the asymptotic
regime, and its value is 4/7, if one assumes that the horizontal profile of the
plume is gaussian.

The maximum number of plumes that may reach the base 
of the solar convection zone, when taking the counterflow into account, is around
1000, which means that the spacing between plumes is about 60 Mm. Note however
that this evaluation is based on a very strong
 assumption: namely that the totality of the convective
energy flux (enthalpy plus kinetic) is carried by the plumes, thus
neglecting the contribution of the interstitial medium.

Finally, under the same assumption, the extent of penetration, at 
the base of the convection zone, depends on the number of
plumes $N$ reaching that level: it varies from $0$
for $N \approx 12000$ to one pressure scale-height for $N \rightarrow 0$.
Unfortunately, the number of plumes cannot be predicted with the simple model
presented above. Let us recall that helioseismology yields a value of about 
$0.1 H_p$, with the assumption that there is sharp transition from adiabatic to
radiative slope at the edge of the penetration layer (Roxburgh \& Vorontsov 1994).
With plumes of different strength, the transition would be smoother and more
difficult to diagnostic.

All these results are independent of the top boundary conditions on the 
mass flux and on the initial momentum flux carried by the plumes. 
What determines the flow is the convective energy flux near the surface,
which would be conserved along the plume if there were no radiative transport.
Unfortunately the simple model presented above is unable to render the complex
dynamics of these layers, where important radiative exchanges occur. They can
only been described by full-fledged 3-dimensional hydrodynamical simulations, such
as performed by \AA. Nordlund, R. Stein and their collaborators. However we cannot
rule out that a sophisticated closure model, well beyond what has been described
above, may succeed in better capturing the main properties of such turbulent
plumes (see V. Canuto's contribution to this workshop).

How do these plumes change our picture of stellar convection zones?
Plumes presumably exert an important effect on the thermal structure -- that is,
on the small deviations from the isentropic stratification which are known to induce
meridional circulation. The reason for this is that the heat transport by the plumes
is of advective nature, whereas in most models so far, as for example in the 
mixing-length treatment, it has been represented by a diffusive process.

They will also play a major role in the dynamics of a convection zone,
and in determining its rotation profile.  Since the ambiant angular momentum will be
entrained into the plumes, these will spin up with depth, and they will concentrate
both vorticity and helicity. Owing to their robustness, such vortices could well be
the cause of the observed differential rotation, especially of its peculiar
property of being quasi-independent of depth (Brown et al. 1989; Kosovichev et
al. 1997). 

Furthermore, one can easily imagine that plumes play a major role in the
generation and in the confinement of magnetic fields. Since they are sites of high
helicity, they will be the seat of strong alpha-effect. The dynamo simulations
of  Brandenburg et al. (1996) have shown that laminar diving plumes are 
indeed playing an important part in the field generation and its storage 
at the base of the convective layer. In more recent simulations, they have been
clearly observed to `pump down' the magnetic field, which is definitely an advective
process (Dorch 1998).

To conclude, we have yet to assess all the consequences of the presence of
plumes in stellar convection zones. For that we have to rely mainly on numerical
simulations, since the stratification is much stronger in stars than in the
Earth atmosphere, and since it cannot be reproduced in the laboratory. These
simulations have not reached yet a sufficient resolution to adequately describe
turbulent plumes, but they are progressing at fast pace, and their results can be
checked through helioseismology.


\end{document}